# Physics for the environment and sustainable development

Juergen Kurths[1] (kurths@pik-potsdam.de)

Ankit Agarwal[1,2] (ankit.agarwal@hy.iitr.ac.in)

Ugur Ozturk[3,4] (ugur.oeztuerk@uni-potsdam.de)

Shubham Sharma[3] (shubham.sharma@gfz-potsdam.de)

Norbert Marwan[1,5] (marwan@pik-potsdam.de)

Deniz Eroglu[6] (deniz.eroglu@khas.edu.tr)

[1] PIK—Potsdam Institute for Climate Impact Research, Member of the Leibniz Association, 14473 Potsdam, Germany

[2] Department of Hydrology, Indian Institute of Technology Roorkee, 247667 Roorkee, India

[3] Helmholtz Centre Potsdam–GFZ German Research Centre for Geosciences, 14473 Potsdam, Germany

[4] Institute of Environmental Science and Geography, University of Potsdam, 14476 Potsdam, Germany

[5] Institute of Geosciences, University of Potsdam, 14476 Potsdam, Germany

[6] Faculty of Engineering and Natural Sciences, Kadir Has University, 34083, Istanbul, Turkey

**Abstract**

A reliable understanding of the Earth system is essential for the life quality of modern society. Natural hazards are the cause of most life and resource losses. The ability to define the conditions for a sustainable development of humankind, to keep the Earth system within the boundaries of habitable states, or to predict critical transitions and events in the dynamics of the Earth system are crucial to mitigate and adapt to Earth system related events and changes (e.g., volcanic eruptions, earthquakes, climate change) and to avert the disastrous consequences of natural hazards. In this chapter, we discuss key concepts from nonlinear physics and show that they enable us to treat challenging problems of Earth sciences which cannot be solved by classic methods. In particular, the concepts of multi-scaling, recurrence, synchronization, and complex networks have become crucial in the very last decades for a substantially more profound understanding of the dynamics of earthquakes, landslides, or (palaeo-)climate. They can even provide a significantly improved prediction of several high-impact extreme events. Additionally, crucial open challenges in the realm of methodological nature and applications to Earth sciences are given.

Keywords: Earth Systems, Nonlinear Physics, Scaling, Climate Hazard

## 1. Introduction

Inventions of the thermoscopes and barometers in the early 17th century enabled studying physical parameters of the climate variables, such as precipitation, temperature, and pressure. Exploring the Earth system in detached disciplinary practices became convenient with these early instruments for limited geographic locations. The disciplinary assessment of the individual Earth system components continues to understand fundamental mechanisms. They have been regarded as autonomous systems in their own right and further broken down into more specialized subsystems. One standard topic is to study, for instance, precipitation concerning more prominent atmospheric modes [1]. Until the last decades, this traditional practice of studying the four major spheres of the Earth system, i.e., the atmosphere, hydrosphere, biosphere, and geosphere, independently continued. However, the Earth behaves as an integrated complex system with nonlinear interactions and feedback loops between and within them [2]. For example, the influence of significant volcanic eruptions on climate oscillations proves a vital link between the geosphere and the atmosphere [3]. The increasing availability of data and the rising concerns related to shifts in the global climate system, concomitant extremes, and natural hazards have urged the development of a more *holistic understanding of the Earth System* in the last decades (Figure 1).

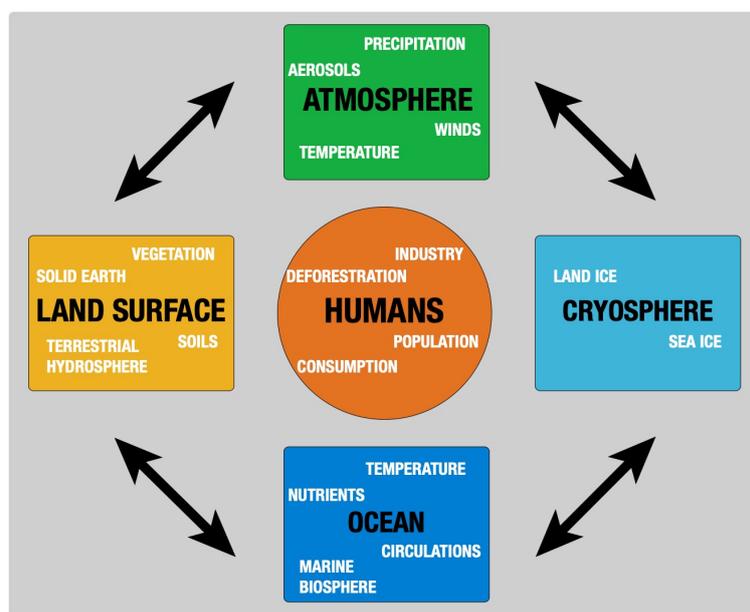

Figure 1: Scheme of rich connections within main components of the Earth system.

Furthermore, it was inherently assumed that various Earth processes are *scale-invariant*, i.e., we can expect a phenomenon to occur in several scales when we observe its occurrence on only one scale [4]. Indeed, the scale-invariance theory was applied to many fields, such as frequency-size distributions of rock fragments, faults, earthquakes, volcanic eruptions, landslides, and oil fields but not all on the Earth Systems. Nevertheless, even the lack of scale invariance means that information is stored and perceived differently at different scales, resulting from mutual interactions of intertwined sub-components interacting over a wide range of scales. Generally, a deep understanding of these multi-component interactions between the different subsystems of the Earth system, including human activities, requires an interdisciplinary approach where concepts from various fields of physics and complex systems science are vital elements [5].

Understanding interacting Earth systems as a giant complex system using only instrumental records is insufficient since such measures cover only a very narrow window of the planet's history. Earth is continuously experiencing natural events such as geological and tectonic processes, climate change, biological and chemical activities. Although the instruments to record such events were not available before the 17th century, fortunately, various natural and complex formations, such as stalagmites, marine and lake sediments, or even trees, have recorded such events in their structures as proxy records. Investigating these archives to reveal the hidden preceding events helps us understand the dynamics and predict the oncoming behavior of the associated natural events on Earth. For this purpose, paleoclimatology, a field of climate science to understand (ancient) climate without direct measurements, has reached a sufficient matureness to reveal significant climate periods, such as glaciations or abrupt global temperature rises, by dating and analyzing the proxies [6].

Whereas the paleoclimate variations as derived from the geoscientific archives are only estimates and contain a degree of uncertainty, the significant climate periods of the driving processes such as the Milankovich cycles can be determined with high accuracy because the equations of motion for the dynamics of the Earth orbit in space can be solved with a reasonable approximation using the Hamiltonian mechanics. However, the celestial sign of objects in the solar system is, in general, a many-body system, where the planets' gravitational fields mutually influence their orbits around the Sun. Solving such a many-body problem (and even of a three-body system) is not simple and was at the forefront of science for a long time [7]. In this spirit and honor of the 60th birthday of the King of Sweden, Oscar II, in 1887, a prize was announced to solve the many-body problem. The French mathematician Henry Poincaré finally won this prize with his seminal work on the three-body system and discovering the chaotic nature of the orbits [8]. In this work, he proved an important theorem that affects the recurring orbits of the interacting objects in a celestial system and is also a fundamental property of many complex dynamical systems: the now well-known recurrence theorem, which states that a (conservative) system recurs infinitely many times as close as one wishes to its initial state. The property of recurrence is not only of the fundamental importance of dynamical systems; it is also a fundamental principle in the Earth sciences at all temporal and spatial scales.

## 2. Nonlinear Concepts

The vigorous progress in exploring nonlinear dynamics in the 1980s and 1990s opened new doors for a more appropriate analysis of complex nonlinear systems, such as lasers, the human brain, power grids, and the Earth system [9]. Techniques for estimating fundamental characteristics of nonlinear systems, such as fractal dimension, Lyapunov exponents, Kolmogorov entropy, and Hurst exponents, were developed and applied to various disciplines [10]. However, these methods are mainly helpful in low-dimensional processes and are not appropriate for understanding the Earth system from data.

Shortly afterward, other essential concepts such as recurrence plots [11,12], synchronization [13], wavelets [14], and complex networks [15] have been developed to explore dynamical and structural properties in high-dimensional spatiotemporal systems. They have been proven to be very promising even for the study of the Earth system. In the following, we shortly describe such basic nonlinear concepts and present some paradigmatic applications in Earth sciences.

## 2.1 Multi-scaling

Various Earth processes are assumed to be *scale-invariant* [4]. An essential law is the size distribution of natural events, meaning that prominent events are less frequent when compared to smaller ones. Deriving an adequate size distribution of natural events would estimate the rarity and likelihood of a specific event. Hence, one major challenge of studying the occurrence, frequency, and intensity of climate-driven natural extremes and natural hazards is these events' spatial and temporal scaling to derive adequate risk estimates. One way to analyze the scaling of natural hazards is to use the size-frequency distribution $p(x)$ (x stands, e.g., for landslide area). For instance, $p(x)$ of landslides follows a power-law probability density function in an area—with arbitrary dimensions—independently of their source mechanism (e.g., earthquake- or rainfall-induced):

$$p(x) = (\alpha - 1)x_{min}^{\alpha-1}x^{-\alpha} \qquad (1)$$

with α the power exponent, valid for $x \geq x_{min}$ [16].

Similarly, the famous Gutenberg-Richter power-law [17,18] scales the seismic activity to assess earthquake hazards for different events magnitudes *m*. It states that earthquake magnitudes *m* are distributed exponentially as

$$\log N_{m \geq M} = a - bM \qquad (2)$$

where $N_{m \geq M}$ is the number of earthquakes with magnitude $m \geq M$, *a* is a constant, and *b* is the scaling parameter. The scaling parameter *b* determines the relative frequency of small and large earthquakes. The estimation of the *b* is around 1.0, with deviations up to 30% in seismically active regions [19]. A real example of this particular case is presented in Section 3.1.

Information of the earth system processes information can be stored and perceived differently at multiple scales. The information observed at one scale often cannot be directly used as information at another one. Scaling approaches address the changes at the measurement scale and plays an essential role in Earth sciences by providing information at the scale of interest.

Determining scaling properties of geophysical variables provides an alternating way to obtain information about the associated process. The processes with similar statistical properties at different scales are said to be self-similar which can be described mathematically as [20]:

$$\phi(x) = \lambda^{-\beta}\phi(\lambda x) \qquad (3)$$

where *x* is the finer spatial resolution (scale), β is the scaling exponent, λ is the ratio of the large resolution, $\lambda x$ to the small resolution *x*, and $\phi$ is the geophysical property or variable of interest. A field is said to be spatially scaling with respect to the moment, *q*, if the following relationship holds [21]:

$$E[(\phi_\lambda)^q] \propto \lambda^{K(q)} E[(\phi_1)^q] \qquad (4)$$

where, $K(q)$ is the scaling exponent associated with the moment of order q. If the exponent $K(q)$ is linear w.r.t $q$, the process has simple scaling. On the other hand, if the scaling exponents, or slopes, are a nonlinear function of $q$, then the process is said to be *multi-scaling*. This concept of scaling and multi-scaling has been used widely in many scientific fields, including hydrology and ecology. For instance, the wavelet analysis can decompose high-resolution non-stationary spatial information into non-stationary fields of increasingly coarser spatial scales [22]. The wavelet and the corresponding scaling function are a function to decompose spatial information into directional components explained by the wavelet coefficients.

**2.2 Recurrence Analysis**

The seminal work of Poincaré in 1890 [8] played a central role in the qualitative theory for nonlinear dynamics (see Section 1). Poincaré presented a method that provides a local and global analysis of nonlinear dynamical systems by the Poincaré recurrence theorem and stability theory for fixed points and periodic orbits. This theoretical finding is compellingly confirmed by the real world, where recurrences can be observed in our daily life and across all scientific disciplines. Therefore, the investigation of recurrences has attracted much attention, and several approaches have been developed for this purpose.

Among the various methods for studying recurring processes, the power spectrum analysis is one of the best known and widely used techniques for identifying periodicities in time series [Schuster 1898]. Wavelet analysis reveals similar information, additionally providing the change of the detected periods over time (see Section 2.1). Coming from the theory of dynamical systems and based on Poincaré's recurrence theorem, the *recurrence plot* (RP) is another fundamental approach that can be used to investigate recurring features in time series and even in spatial data [11,12]. In a given $m$-dimensional phase space, neighboring two points are called recurrent if the distance between their state vectors is closer than the threshold $\varepsilon$. Formally, for a given trajectory $\boldsymbol{x}_i$ ($i = 1, \dots, N, \boldsymbol{x} \in R^m$), the recurrence matrix $\boldsymbol{R}$ is defined as

$$R_{i,j} = \begin{cases} 1, & \text{if } \|\boldsymbol{x}_i - \boldsymbol{x}_j\| \leq \varepsilon \\ 0, & \text{otherwise} \end{cases} \tag{5}$$

where $\|\cdot\|$ is a norm of the adopted phase space. The graphical representation of the recurrence matrix $\boldsymbol{R}$ is the RP (Figure 2). RP of different dynamical behavior represents different particular features (Figure 2). Such differences can be quantified with the measures of recurrence quantification analysis such as determinism (the fraction of recurrence points that form diagonal lines in the recurrence plot), laminarity (the fraction of recurrence points that form vertical lines), and recurrence rate (the percentage of recurrence points in a recurrence plot). These measures are used to find changes in the dynamics of a process (e.g., in climate), to classify the dynamics (e.g., random, chaotic, regular) [12], or to identify interrelationships and coupling directions in coupled systems [23].

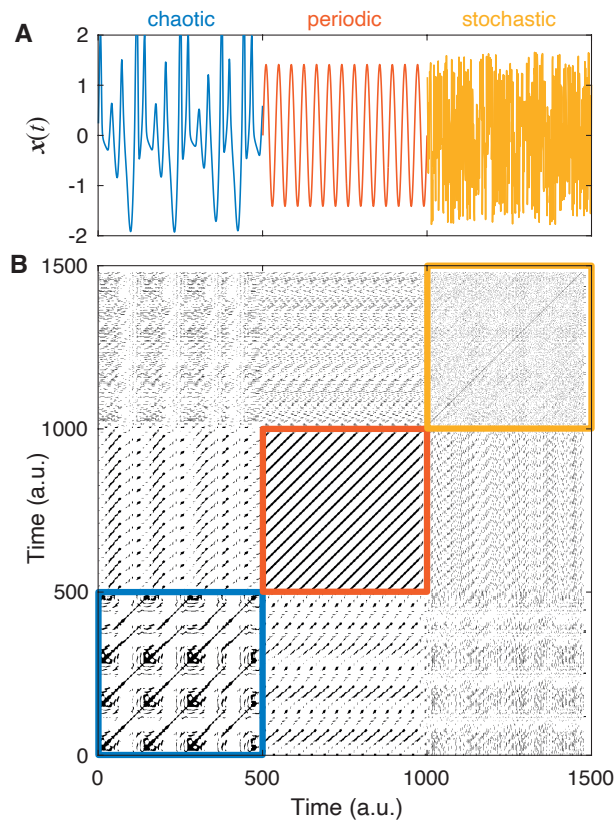

Figure 2: (A) Time series representing switching between different dynamical regimes, from chaotic via periodic to stochastic, each lasting 500-time steps. (B) A recurrence plot (RP) is representing the recurrence of a state at a given point in time (*x*-axis) at another point in time (*y*-axis). Different dynamics cause typical recurrence patterns, which can be used to detect these changing dynamical behaviours. Continuous long diagonal lines in the RP indicate the periodic window, shorter diagonals show the chaos, and single points appear in the stochastic part.

## 2.3 Complex networks and event synchronization

Essential challenges in climatology are quantifying the spatial extent of climate extremes and early forecasting procedures of their dynamical behavior. Such forecasting relies predominantly on numerical models which solve physics-based coupled systems of partial differential equations. Starting with Richardson in the 1920s, it has been a long way to the first successful prediction in 1950 and eventually to today's highly sophisticated general circulation and Earth system models. Despite multiple efforts of these methods, their predictive power, especially for extreme events, can be rather limited. A primary reason for this is that in particular long-range interactions, called *teleconnections*, and their interaction with more regional interactions may not be well represented or even absent in such models.

Therefore, a quite different approach has been suggested, a network-based presentation of climate phenomena, called *climate networks*. The main idea is to get additional information by capturing the evolving interactions of different locations, regarded as nodes, through similarity measures, such as the Pearson correlation, mutual information, or the Granger causality, from spatio-temporal observational data. An important description of such similarity of strong events is the *event synchronization* approach [24], inspired by Christiaan Huygens' detection of synchronization in the 17th century. Here we consider the occurrence of extreme events, e.g., rainfall, in a synchronized manner at different locations, even far away ones.

The final complex network is then represented by an adjacency matrix $A$, which encodes the links between the nodes $i$ and $j$ as follows:

$$A_{i,j} = \begin{cases} \text{nonzero}, & \text{if variability at node } j \text{ is similar (or synchronized) to node } i \\ 0, & \text{otherwise} \end{cases} \quad (6)$$

The value of the elements of $A$ represents the weight of the link obtained from quantifying similarity (Figure 3).

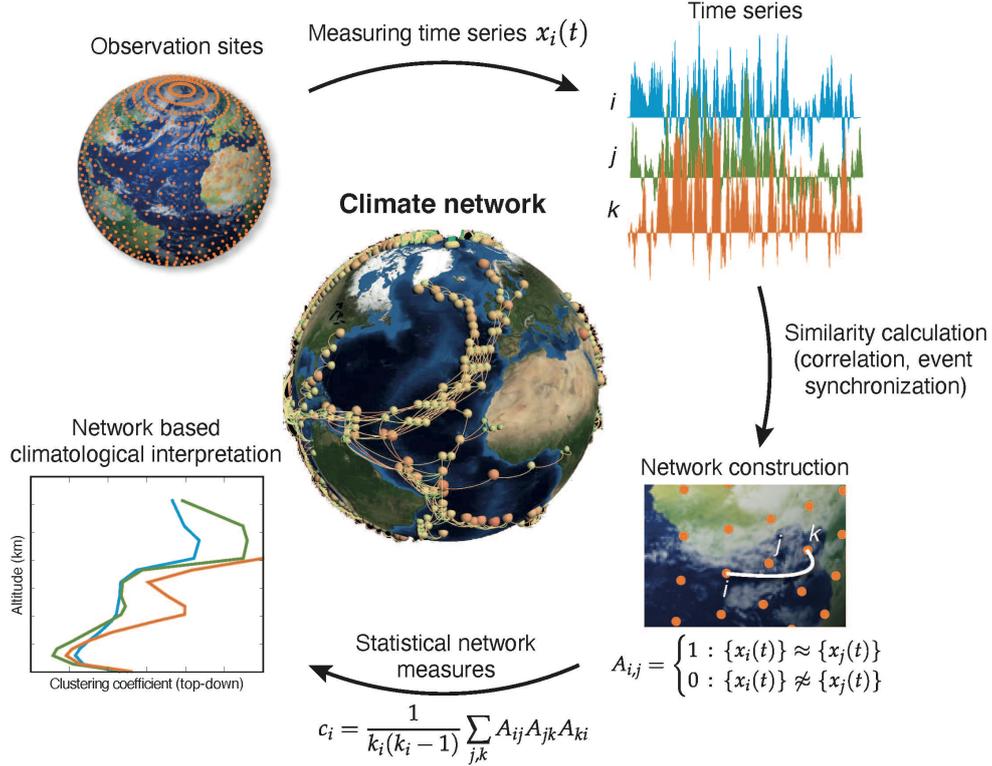

Figure 3: The climate-network framework as a tool for prediction. Observational data of physical quantities, e.g., temperatures, are available at different geographical locations. These data can be used directly or via a reanalysis (numerical weather model) which assimilates and maps them onto a regular grid. Thus, a time series of the regarded physical quantity is available for each climate network node (observational site or reanalysis grid point). Cooperativity between nodes can be detected from the similarity in the evolution of these time series and translated into links connecting the corresponding nodes. The links or their strengths may change with time. These nodes and their links constitute the evolving climate network represented by the adjacency (connectivity) matrix $A$ (Eq. 6). The analysis of this network can enable early predictions of climate phenomena and provide insights into the physical processes of the Earth system.

There are various generalizations of this construction, particularly to emphasize multilayer networks, which enable variables from different subsystems.

The reconstructed adjacency matrix $A$ allows us to calculate standard network measures such as degrees, clustering coefficients, or betweenness but also to identify teleconnections. It has been shown recently that climate networks provide ideal tools for exploring even large climate data to uncover spatiotemporal patterns leading to new physical insights into the climate system [1]. Moreover, they have a strong predictive potential, i.e., they enable the development of new forecasting methods. Examples of up-and-coming applications are given in sections 3.3, 3.4, 3.5.

## 3. Applications of Nonlinear Dynamics in Earth System

Vigorous progress in nonlinear science contributed to detecting, attributing, and understanding the Earth system, reducing uncertainties, and projecting future climate changes. In this section, we discuss some significant contributions of nonlinear physics in Earth system sciences.

### 3.1 Earthquakes and the Gutenberg–Richter law

A proper fitting of the power law is essential to study most natural hazards, particularly earthquakes (Eq. 1). The Gutenberg–Richter law (Eq. 2) represents scaling in earthquakes, as power-law distribution makes it scale-invariant. An example of the scale parameter $b$ for central California for 20 years (2001-2020) is illustrated in Figure 4. The California region has a $b$ of 1.0, which is as per the global average, meaning that central California has the same relative frequency of small and large earthquakes. However, the magnitude threshold parameter $M_c$ must be selectively applied above crossover magnitude for larger earthquakes with significant seismic moments [25]. The Gutenberg–Richter law accurately describes the shallow seismicity. However, it is not the only scaling law for all levels of earthquake events; the distribution of deeper earthquakes was observed as following a bimodal (multi-scaling) pattern [26].

It is also crucial to accurately estimate scaling parameter $b$ (Eq. 2) from the earthquake events to characterize the seismicity activity (see Section 2.1) sensitively. There is an inverse correlation between $b$ and the differential stress, which was revolutionary in that $b$ can act as an indicator of stress accumulated around the fault volume [27]. This observation is used in the study done before and after the vast 2011 Tohoku-Oki earthquake with a high slip area, where an increase in $b$ is observed as a large amount of stress was released [28]. Another employment of this observation is studying the structural anomalies in the crust and identifying the volumes of magma in an active volcano. A study performed at two active volcanoes [29], Mt. St. Helens and Mt. Spurr, shows a relatively high $b$ ($\geq 1.3$) due to the presence of material heterogeneity and high thermal gradient. This high $b$ is why these volcanoes are less likely to host large earthquakes but frequent small ones. A typical intraplate $b$ is around 0.8, making intraplate regions prone to large earthquakes over a short recurrence time. However, the scaling parameter $b$ is not the perfect parameter to measure seismicity at all magnitude scales. The tail of the $\log(N_m > M)$ vs. $M$ relation holds for only a certain range of magnitudes. A nonlinear fit is a better approximation for smaller ($M_c \leq 3.4$) and larger ($M_f \gtrsim 7$) magnitudes. A reason, for the deviation from the power-law for smaller earthquakes than $M_c \leq 3.4$ (Figure 4), is the incompleteness of catalogs. For large earthquakes, a reason is the saturation of the magnitude scale and the long recurrence time which makes them missing from the catalogues because they are often too short.

**High scaling parameter *b* indicates a lower chance of observing significant seismicity while the frequency of small earthquakes is high. However, smaller magnitude events are observed way less than the indicated by b due to insufficient seismic network coverage.**

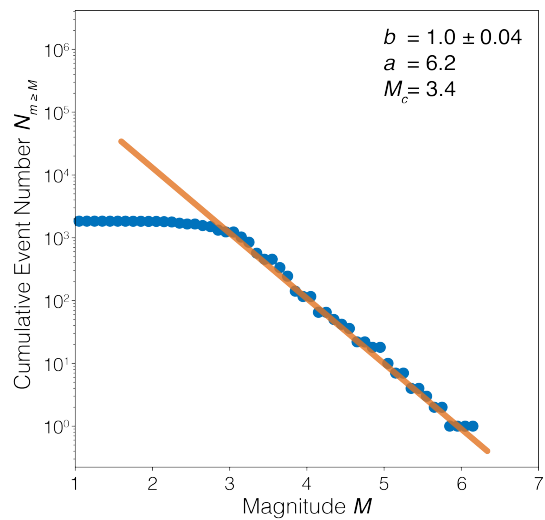

Figure 4: Frequency magnitude distribution (FMD) for earthquakes in central California between 2001 and 2020. The red line shows a fit to the cumulative frequency and has a slope (*b*-value) of 1.0. The magnitude cut-off, $M_c = 3.4$, is used for estimating the scaling parameter b.

### 3.2 Recurrence Plot Application

Recurrence is a fundamental principle in the Earth sciences at all temporal and spatial scales: starting, e.g., from the key principle of the "doctrine of uniformity", over the rock cycle, glaciation cycles, active geysers, to alternating sediment layers (to mention only a few). One crucial phenomenon with complex recurrence patterns is the climate. Among others, a primary driver of climate is solar insolation, modulated by mutual variations of the Earth's orbit around the Sun and the tilt of the Earth axis, responsible for seasons, changes in global temperature, and glaciations. This influence has been discovered already in the first half of the last century by investigating annually layered lake sediments [30], and considering the Earth's orbital parameters by Milankovich [31].

Recurrence plots (see Section 2.2) provide a powerful framework to study the dynamics of the climate by their recurrence properties. As an application, the dynamics of the Cenozoic climate will be investigated by recurrences properties in a selected palaeoclimate proxy record. Such studies are essential to advance our understanding of the past and aim to improve climate models to better forecast future climate change and its impacts, as well as increase our understanding of climate dynamics.

Calcareous lake sediments, speleothems, or benthic foraminifera stores environmental conditions by changing their geochemical and petrographic composition. The study of stable isotopes is an active field to derive past environmental and climatic conditions. For example, the temperature-dependent fractionation of oxygen isotopes is the key to reconstruct global seawater temperatures and ocean circulation by using planktonic and benthic foraminifera. Ongoing deep ocean drilling programs and novel quantitative methods like clumped isotope thermometry provide new insights with improved quantification, increasing temporal resolution, and ever-smaller time uncertainties. The recently developed temperature reference curve for the Cenozoic [32] is an example with a temporal resolution of up to 2000 years and covering 66 million years. This period is crucial because it provides an analog of future greenhouse climate and how (and which) regime shifts in large-scale atmospheric and ocean circulation can be expected in a warming world [33]. The outstanding high resolution of this record allows to study and compare recurrence properties of selected time intervals. The recurrence plot indicates the different climate regimes of hothouse, warmhouse, coolhouse, and icehouse by their very distinct recurrence pattern (Figure 5). During the Miocene (18 to 14 Ma ago), the climate was in a warmer state more similar to the warmhouse than the coolhouse, visible by some recurrences linking this period to the late Eocene. The fine-scale pattern of the recurrence plot reveals more details, such as the change from the 41 ka cycles to 100 ka Milankovich cycles of glaciation during the mid-Pleistocene transition.

**Recurrence analysis of climate time series indicates different dynamical regimes, such as chaotic or predictable dynamics, thus, detects critical transitions between different climate periods.**

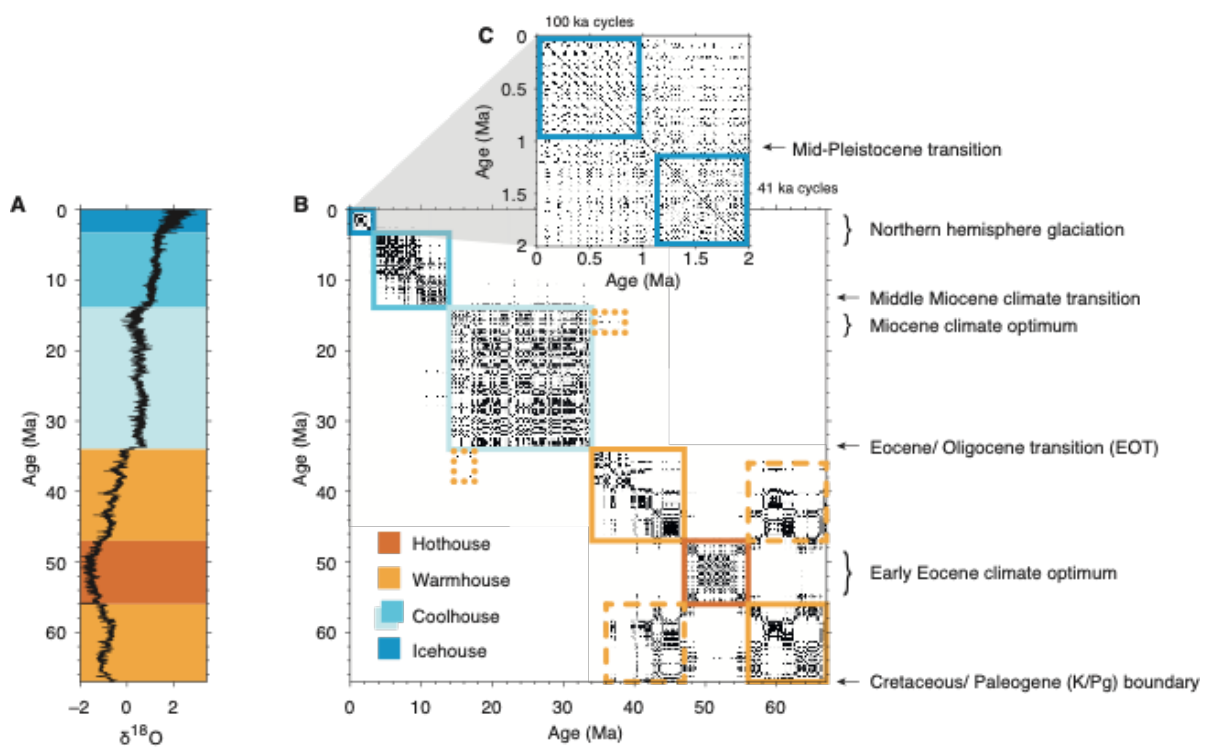

Figure 5: RP of a palaeoclimate time series. (A) Palaeoclimate variation indicated by Oxygen isotope measurements from marine sediments (CENOGRID). Lower values correspond to a warmer global climate. (B) The RP indicates the different climate regimes of hothouse, warmhouse, coolhouse, and icehouse by their very distinct recurrence pattern. During the Miocene (18 to 14 Ma ago), the climate was warmer state more similar to the warmhouse than the coolhouse, visible by some recurrences linking this period to the late Eocene (marked by the dotted box). (C) The fine-scale pattern of the RP reveals more details, such as the change from the 41 ka cycles to 100 ka cycles of glaciation during the mid-Pleistocene transition.

### 3.3 Extreme Rainfall Teleconnections and Monsoon Prediction

The Indian summer monsoon is an intense rainy season lasting from June to October. The monsoon delivers more than 70% of the country's annual rainfall, which is India's primary source of freshwater. Although the rainy season happens every year, the monsoon onset and withdrawal dates vary within a month from year to year. Such variability strongly affects the life and property of more than a billion people in India, especially those living in rural areas and working in the agricultural sector, which employs 70% of the entire population. So far, only Kerala in South India receives an official monsoon forecast two weeks in advance, while the other 28 states rely on the operational weather forecast of about five days [34]. A much better forecast has been recently reached by combining two nonlinear concepts: complex climate networks and a tipping-element approach.

In the first step, from rainfall data from the Asian Precipitation Highly Resolved Observational Data Integration Towards the Evaluation of Water Resources (APHRODITE) and the high-resolution satellite product Tropical Rainfall Measurement Mission (TRMM) 3B42 complex networks were retrieved via the event synchronization technique (see Section 2.3). This exploratory network-based analysis of extreme rainfall across the Indian subcontinent enabled for the first time the identification of critical geographical domains displaying far-reaching links, influencing distant grid points [35]. In particular, North Pakistan and the Eastern Ghats turn out to be crucial for the transport of precipitation across the subcontinent.

In the second step, a tipping-elements approach of the measured daily mean air temperature and the relative humidity at these two sensitive regions allowed us to uncover the critical nature of the spatiotemporal transition to the monsoon. It was especially found that the temporal evolution of the daily mean air temperature and the relative humidity exhibits critical thresholds on the eve of the monsoon. A highly developed instability occurring in these regions creates the conditions necessary for spatially organized and temporally sustained monsoon rainfall.

Based on this knowledge, a scheme was developed for forecasting the upcoming monsoon onset in the central part of India 40 days in advance, thus considerably improving the time horizon of conventional forecasts. The new scheme has proven its skill (73% of onset predictions correct) not only in retrospective (for the years 1951-2015) but showed to be successful in predicting future monsoons already five years in a row since its introduction in 2016. The methodology appears to be robust under climate change and has proven its skill also under the extreme conditions of 2016, 2018, and 2019.

Further successful applications of this network-based concept are El Niño forecasts beyond the spring barrier, predicting droughts in the central Amazon 12 to 18 months in advance, or forecasting extreme rainfall in the Eastern Central Andes [36].

**A network-based analysis of climate data can provide predictive power for mitigating the global-warming crisis and societal challenges.**

### 3.4 Understanding landslide distributions

As explained in Section 2.1, successfully fitting a global power-law distribution (Eq. 1) to landslides would help us to understand whether we lack information in hazard and risk models. Although the distribution of spatial landslides follows a power-law distribution. Just as the case of Gutenberg Richter, the power exponent is valid until a minimum value (Eq. 1) [16], the rollover below the minimum is found in two different forms (i) the double Pareto distribution and (ii) the inverse Gamma distribution according to different studies [37,38]. Like other universal scaling laws [39], it is also expected to have a universal power exponent for the landslide events. However, a lack of data makes studying the problem impossible in a better resolution especially at the function's tail. [37,38]. Like other universal scaling laws [39], it is also expected to have a universal power-law distribution for the landslide events. However, a lack of data makes studying the problem impossible in a better resolution. Most studies rely primarily on landslide inventories collected following a significant landslide triggering event, such as the 1994 Northridge earthquake ($M_W$ 6.4). Landslides have also been found to exhibit temporal scaling or clustering besides spatial and geometric ones. Although some studies suggest a global power exponent $\alpha = 2.3\pm0.6$, the physical process is unknown to implement a functional probabilistic multi-hazard assessment [40].

Besides the power-law-based approximation models, ample practices offer linear solutions to study natural hazards, making a nonlinear application redundant. An example would be Newmark's sliding block analysis. It estimates the displacement potential of hillslopes under seismic loading (i.e., acceleration). This hypothetical displacement aims to indicate the likelihood of failure under seismic loading as a function of hillslope inclination and seismic acceleration. For example, landslides related to the 2016 Kumamoto earthquake ($M_W$ 7.1) caused significant damage, especially to the infrastructure, such as highways (Figure 6A). Although landslide locations correlate well with the seismic waveforms based on a physics-based ground-motion model [41], the Newmark's distances highlight particularly elevated gradients in the landscape (Figure 6B).

Rainfall rather decreases the slope stability by altering cohesion, elevating the landslide susceptibility in most cases. In some other cases, rainfall could also mobilize the superficial surface material leading to the debris flows. However, unlike an earthquake, rainfall is not introducing a direct force on the hillslopes to estimate rainfall impact on landslides. Hence, most of the time, statistical methods are applied to forecast rainfall-induced landslides. One standard tool is to use statistically derived rainfall intensity-duration thresholds above which landslides are triggered. The logic behind this is that high-intensity rainfall triggers landslides and moderate intensity, but long-duration events would increase the landslide susceptibility. Therefore, several spatial classification models are developed trying to relate landslide activity to rainfall distribution.

Another notorious example is that the extreme rainfall flux over a region during tropical storms alike mechanisms—as previously explained—might already highlight the landslide-prone regions on large spatial scales. It is possible to estimate/cluster the rainfall motion over large areas, countries, or continental scale by blending event synchronization and complex network methods (see Section 2.3). These results can help track landslide activity along the path of extreme rainfall. As an application, the extreme rainfall trajectories over the Japanese archipelago were estimated using event-synchronization [42]. The density of extreme rainfall tracks aligns well with the landslide distribution (Figure 6C–E).

**The power-law distribution can model landslide distributions, and using nonlinear methods such as event-synchronization, it is possible to describe spatial landslide distributions as a function of rainfall distributions.**

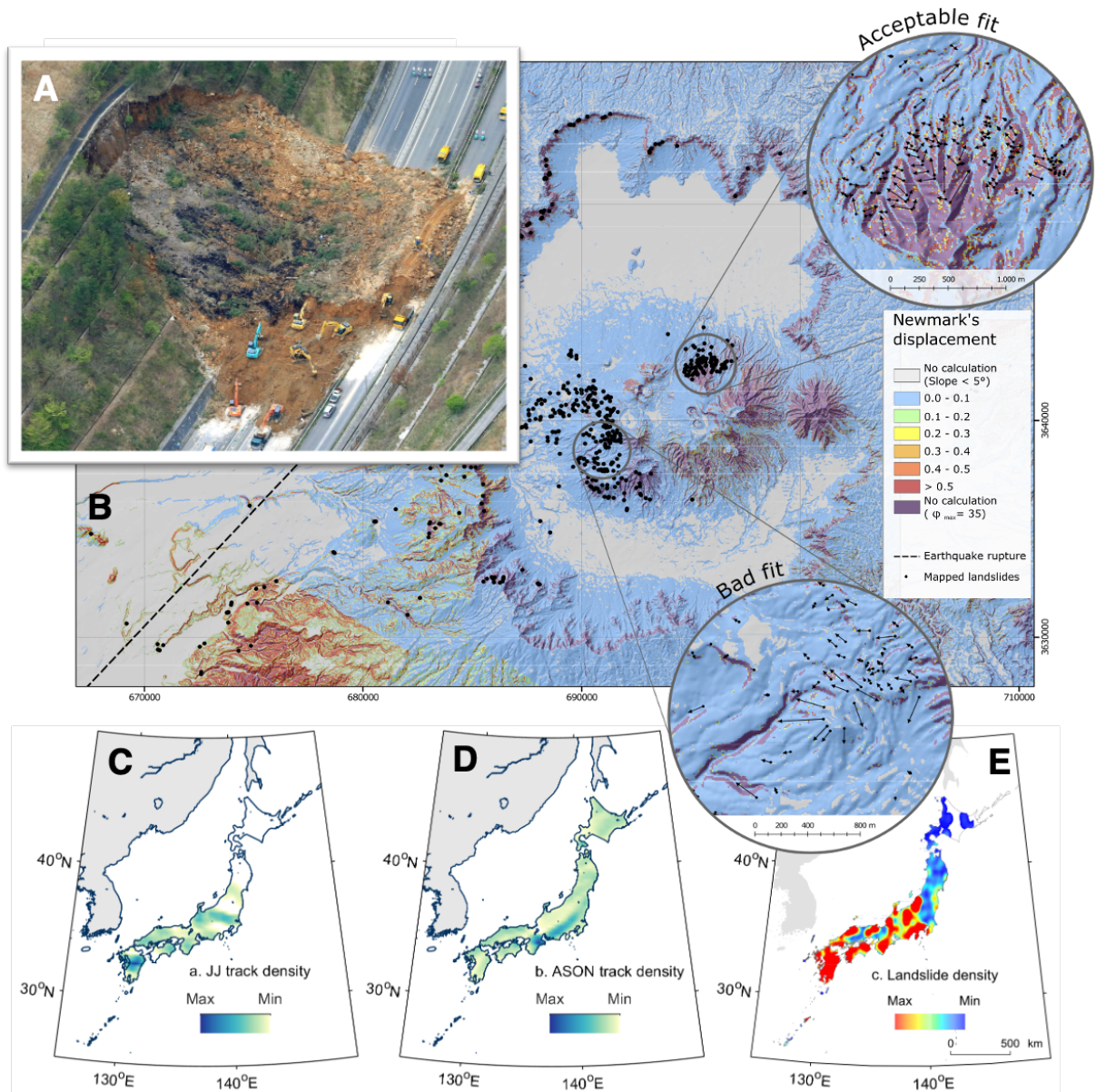

Figure 6: (A) Example of a cut slope failure by the Oita Expressway following the 2016 Kumamoto earthquake ($M_W$ 7.1), the photo is taken from Dave Petley's landslide blog (https://blogs.agu.org/landslideblog/2016/04/18/kumamoto-Earthquake-1/). (B) Newmark's displacement of the 2016 Kumamoto earthquake ($M_W$ 7.1) in Kyushu, Japan (UTM-52). In certain regions, the elevated displacement correlates well with the mapped landslides, while in some others, it is relatively poor. The concentration of extreme precipitation streamlines during (C) June and July (JJ), and (D) August to November (ASON), normalized by cumulative above 95% extreme rainfall for the same period between 1998 and 2015 based on TRMM (Tropical Rainfall Measurement Mission) rainfall estimates. (E) Normalized rainfall-triggered spatial landslide density-weighted by log-transformed landslide volumes calculated from an inventory of 4744 events and smoothed by kernel density estimation onto a 5×5 km grid by [43]; white areas have no data.

### 3.5 Multi-scale Sea-Surface Temperature (SST)

Climatic systems are complex systems comprised of multiple feedbacks and interactions. In such systems, the coupling between climate variables takes place at different time and spatial scales. Untangling this multi-scale variability and interactions of a climatic process are vital as they would improve the understanding of global climate and its variability. Hence climate networks are constructed (section 2.3) at different time scales considering each SST grid cell as a node, and edges are created between all pairs of nodes based on statistical relationships. First, SST data is decomposed at different time scales using wavelet (section 2.1), and then the Pearson correlation between all pairs of nodes is calculated at a corresponding time scale. Finally, significance-based pruning is applied to retain only highly correlated edges in the network. The network is constructed by applying a 5% link density threshold, which is well-accepted for the network construction. Multiple testing was employed to avoid false links.

The network visualization of the original SST data (all scales) reveals short-range and long-range connections between various regions of the Earth. As at a finer scale, there is no significant correlation, and that is expected since we have removed the annual cycle using anomalies. Interestingly, at 8-16 months, we observe mainly two zones with many significant correlations in the equatorial Pacific and Indian Ocean dipole, which are known to impact each other via the atmosphere (Figure 7a). On the next period of 32-64 months, these patterns become more prominent as known ENSO events act on scale up to 2 years (Figure 7b). There is a link between SST in the Southern Ocean to ENSO events via the Southern annual mode, i.e., the north-south movement of the westerly wind belt that circles Antarctica (Figure 7b). The 3-dimensional visualization (Figure 7c) shows several links from the North Atlantic to the South Atlantic. This negative correlation likely exhibits the see-saw response due to the transport of heat from the Southern Ocean to the North Atlantic via the Atlantic Meridional Overturning Circulation (AMOC) [44]. If the AMOC is stronger than before 2000 (as it has been in the period after the year 2000 compared to the years before [45]), more heat is transported towards the North, which leads to a cooling in the Southern Ocean and warming in the subpolar North Atlantic.

**Multi-scale analysis of climatic processes helps to uncover the time-scales of interaction and feedbacks in the climate system that may be missed when processes are analyzed at one timescale only.**

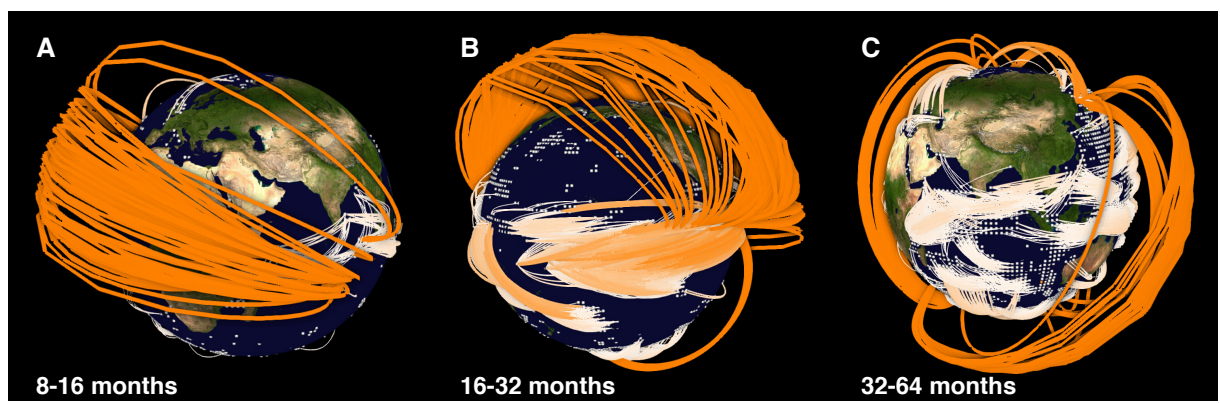

Figure 7: Spherical three-dimensional globe representation of the long-range teleconnections at different timescales in sea surface temperature network [46]. Edge color represents the geographical lengths.

## 4. Outlook

We have shown that basic concepts of nonlinear physics and complex systems science have a strong potential for treating important problems in Earth systems sciences. We have argued that they complement established concepts with new possibilities to reveal entire causal chains of complex phenomena in the Earth system, primarily to reveal new precursor processes of extreme events.

However, it is essential to emphasize that these interdisciplinary approaches are in their infancy and the subject of ongoing research. There are various open challenges in the realm of methodological nature and applications. In the following, some of them are summarized:

- There is a growing recognition in the scientific community and, more broadly, that the Earth functions have to be regarded as an interconnected complex system with properties and behavior characteristics of the system as a whole. These include tipping points, critical thresholds, "switch" or "control" points, strong nonlinearities, teleconnections, chaotic elements, and uncertainties of different origins. Understanding the components of the Earth system is important; however, it is insufficient to understand the functioning of the whole Earth system. Humans are now a significant force in the Earth system, altering key process rates and absorbing global environmental changes. Human activities' environmental significance is so profound that the current geological era is called the Anthropocene [47]. Therefore, there is a strong need to develop a complex global model involving Earth system dynamics, human activities, and environmental boundaries to systematically study the planetary boundaries and tipping points and uncover fundamental principles.
- An important task is to improve our capabilities regarding data-driven inference of governing principles to reach a deeper understanding of the connection between the microscopic dynamics of Earth systems constituents and their nonlinear interactions on the one hand and the dynamics emerging from these interactions at the macroscopic level, on the other hand.
- Combining traditional physics-based modeling and statistical approaches with state-of-the-art machine learning (ML) techniques is necessary to efficiently include the huge amount of available data in a model. However, we would like to emphasize that neither an ML-only nor a scientific knowledge-only approach is sufficient for complex Earth system applications. Hence, we must explore the continuum between mechanistic and ML models, where both scientific knowledge and data are integrated synergistically [48,49]. This approach has picked up momentum just in the last few years [49] and is being pursued in Earth systems [50], climate science [51], and hydrology [52].
- A key driver of further advances is the desire to improve predictions of the behavior of complex systems and especially—for example, in the context of the ongoing global warming driven by the anthropogenic release of greenhouse gases—of the response of complex systems to time-varying external forcing.
- The study of surface processes with nonlinear tools is still not common. Combining nonlinear approaches with linear methods could advance the existing forecasting schemes, especially predicting extreme events. The European floods in summer 2021, which claimed nearly 200 lives

- in Germany alone, are a matchless example that emphasizes that more effort has to be placed to forecast extreme incidents to prevent life loss.
- Climate-driven hazards are rarely an output of a single system. Many of those are in the form of hazard cascades, for example, extreme rainfall initiating a flash flood; high waters lead to carving the river banks and triggering landslides; dislocated loose landslide mass mixes up with the high waters and is transported downstream as a debris flow. However, most of the research that links urban interaction with climate-driven natural hazards are empirical studies. Only recently, the first numerical model (CHASM) is set to describe the informal housing-related changes in the topography and link it to the occurrence rates of landslides [53]. CHASM alike models could connect different earth systems. Blending such physics-based models and a nonlinear causation metric such as event synchronization as a preceding step could enhance our capacity to forecast/predict extreme rainfall-driven natural hazards, such as landslides.
- The simultaneous occurrence of two or more natural extremes impacts the society much stronger than their univariate counterparts are [54]. For instance, a hazard resulting from a dry and hot summer co-occurring is higher than a univariate drought extreme, given it triggers a severe impact, such as a reduction in agricultural productivity, irretrievable loss to property and health, damage to natural ecosystems and public infrastructure. These manifold extremes are called compound extremes/compound events [55]. The investigation on compound extremes has received less attention so far; nevertheless, it has gained significant momentum later across the globe [54,56,57].
- Overall transient central components of Earth systems such as temperature, rainfall, and their control on other processes, such as concomitant natural hazards of droughts or landslides, should be emphasized and studied using more recent comprehensive data.

**Acknowledgements**

UO has been supported by research focus point Earth and Environmental Systems of the University of Potsdam. UO and AA acknowledge Co-PREPARE of DAAD (DIP Project No. 57553291). SS acknowledges support from the DFG research training group "Natural Hazards and Risks in a Changing World" (grant no. GRK 2043/1). DE acknowledges TÜBİTAK (Grant No. 118C236). We thank Georg Veh for helping with visualization.

**References**

[1] Boers N, Goswami B, Rheinwalt A, Bookhagen B, Hoskins B and Kurths J 2019 Complex networks reveal global pattern of extreme-rainfall teleconnections *Nature* **566** 373–7

[2] Steffen W, Sanderson R A, Tyson P D, Jäger J, Matson P A, Moore III B, Oldfield F, Richardson K, Schellnhuber H-J and Turner B L 2006 *Global change and the earth system: a planet under pressure* (Springer Science & Business Media)

[3] Mann M E, Steinman B A, Brouillette D J and Miller S K 2021 Multidecadal climate oscillations during the past millennium driven by volcanic forcing *Science* **371** 1014–9

[4] Ge Y, Jin Y, Stein A, Chen Y, Wang J, Wang J, Cheng Q, Bai H, Liu M and Atkinson P M 2019 Principles and methods of scaling geospatial Earth science data *Earth-Science Reviews* **197** 102897

[5]     Fan J, Meng J, Ludescher J, Chen X, Ashkenazy Y, Kurths J, Havlin S and Schellnhuber H J 2020 Statistical physics approaches to the complex Earth system *Physics Reports*

[6]     Bradley R S 2015 *Paleoclimatology: reconstructing climates of the quaternary*

[7]     Muller R A and MacDonald G J 2002 *Ice ages and astronomical causes: data, spectral analysis and mechanisms* (Springer Science & Business Media)

[8]     Poincaré H 1890 Sur le problème des trois corps et les équations de la dynamique *Acta mathematica* **13** A3–270

[9]     Nicolis G and Nicolis C 2012 *Foundations of complex systems: emergence, information and predicition* (World Scientific)

[10]    Kantz H and Schreiber T 2004 *Nonlinear time series analysis* vol 7 (Cambridge university press)

[11]    Eckmann J 1987 Recurrence plots of dynamical systems *Europhysics Letters* **5** 973–7

[12]    Marwan N, Romano M C, Thiel M and Kurths J 2007 Recurrence plots for the analysis of complex systems *Physics Reports* **438** 237–329

[13]    Pikovskij A, Rosenblum M and Kurths J 2007 *Synchronization: a universal concept in nonlinear sciences* (Cambridge: Cambridge Univ. Press)

[14]    Torrence C and Compo G P 1998 A Practical Guide to Wavelet Analysis *Bulletin of the American Meteorological Society* **79** 61–78

[15]    Newman M E J 2010 *Networks: an introduction* (Oxford ; New York: Oxford University Press)

[16]    Jafarimanesh A, Mignan A and Danciu L 2018 Origin of the power-law exponent in the landslide frequency-size distribution *Natural Hazards and Earth System Sciences Discussions* 1–28

[17]    Gutenberg B and Richter C F 1944 Frequency of earthquakes in California *Bulletin of the Seismological society of America* **34** 185–8

[18]    Gutenberg B and Richter C F 1956 Earthquake magnitude, intensity, energy, and acceleration: (Second paper) *Bulletin of the seismological society of America* **46** 105–45

[19]    Frohlich C and Davis S D 1993 Teleseismic b values; or, much ado about 1.0 *Journal of Geophysical Research: Solid Earth* **98** 631–44

[20]    Blöschl G 1996 *Scale and scaling in hydrology* (Techn. Univ., Inst. f. Hydraulik, Gewässerkunde u. Wasserwirtschaft)

[21]    Peters-Lidard C D, Pan F and Wood E F 2001 A re-examination of modeled and measured soil moisture spatial variability and its implications for land surface modeling *Advances in Water Resources* **24** 1069–83

[22]    Kumar P and Foufoula-Georgiou E 1993 A multicomponent decomposition of spatial rainfall fields: 1. Segregation of large- and small-scale features using wavelet transforms *Water Resources Research* **29** 2515–32

[23]    Eroglu D, McRobie F H, Ozken I, Stemler T, Wyrwoll K-H, Breitenbach S F, Marwan N and Kurths J 2016 See–saw relationship of the Holocene East Asian–Australian summer monsoon *Nature communications* **7** 1–7

[24]    Malik N, Marwan N and Kurths J 2010 Spatial structures and directionalities in Monsoonal precipitation over South Asia *Nonlinear Processes in Geophysics* **17** 371–81

[25]    Pacheco J F, Scholz C H and Sykes L R 1992 Changes in frequency–size relationship from small to large earthquakes *Nature* **355** 71–3

[26]   Zhan Z 2017 Gutenberg–Richter law for deep earthquakes revisited: A dual-mechanism hypothesis *Earth and Planetary Science Letters* **461** 1–7

[27]   Scholz C H 1968 The frequency-magnitude relation of microfracturing in rock and its relation to earthquakes *Bulletin of the seismological society of America* **58** 399–415

[28]   Tormann T, Enescu B, Woessner J and Wiemer S 2015 Randomness of megathrust earthquakes implied by rapid stress recovery after the Japan earthquake *Nature Geoscience* **8** 152–8

[29]   Wiemer S and McNutt S R 1997 Variations in the frequency-magnitude distribution with depth in two volcanic areas: Mount St. Helens, Washington, and Mt. Spurr, Alaska *Geophysical research letters* **24** 189–92

[30]   De Geer G 1926 On the solar curve: as dating the ice age, the New York Moraine, and Niagara Falls through the Swedish Timescale *Geografiska Annaler* **8** 253–83

[31]   Berger A L 1978 Long-Term Variations of Caloric Insolation Resulting from the Earth's Orbital Elements1 *Quaternary research* **9** 139–67

[32]   Westerhold T, Marwan N, Drury A J, Liebrand D, Agnini C, Anagnostou E, Barnet J S, Bohaty S M, De Vleeschouwer D and Florindo F 2020 An astronomically dated record of Earth's climate and its predictability over the last 66 million years *Science* **369** 1383–7

[33]   Burke K D, Williams J W, Chandler M A, Haywood A M, Lunt D J and Otto-Bliesner B L 2018 Pliocene and Eocene provide best analogs for near-future climates *Proceedings of the National Academy of Sciences* **115** 13288–93

[34]   Pai D S and Nair R M 2009 Summer monsoon onset over Kerala: New definition and prediction *Journal of Earth System Science* **118** 123–35

[35]   Stolbova V, Surovyatkina E, Bookhagen B and Kurths J 2016 Tipping elements of the Indian monsoon: Prediction of onset and withdrawal: TIPPING ELEMENTS OF MONSOON *Geophysical Research Letters* **43** 3982–90

[36]   Ludescher J, Martin M, Boers N, Ciemer C, Fan J, Havlin S, Kretschmer M, Kuths J, Runge J, Stolbova V, Surovyatkina E and Schellnhuber H in press Network-based forecasting of climate phenomena *Proceedings of the National Academy of Sciences*

[37]   Stark C P and Hovius N 2001 The characterization of landslide size distributions *Geophysical research letters* **28** 1091–4

[38]   Malamud B D, Turcotte D L, Guzzetti F and Reichenbach P 2004 Landslides, earthquakes, and erosion *Earth and Planetary Science Letters* **229** 45–59

[39]   West G B 2017 *Scale: the universal laws of growth, innovation, sustainability, and the pace of life in organisms, cities, economies, and companies* (Penguin)

[40]   Liu Z, Nadim F, Garcia-Aristizabal A, Mignan A, Fleming K and Luna B Q 2015 A three-level framework for multi-risk assessment *Georisk: Assessment and management of risk for engineered systems and geohazards* **9** 59–74

[41]   von Specht S, Ozturk U, Veh G, Cotton F and Korup O 2019 Effects of finite source rupture on landslide triggering: the 2016 MW 7.1 Kumamoto earthquake *Solid Earth* **10** 463–86

[42]   Ozturk U, Marwan N, Korup O, Saito H, Agarwal A, Grossman M J, Zaiki M and Kurths J 2018 Complex networks for tracking extreme rainfall during typhoons *Chaos* **28** 075301

[43]   Saito H, Korup O, Uchida T, Hayashi S and Oguchi T 2014 Rainfall conditions, typhoon frequency, and contemporary landslide erosion in Japan *Geology* **42** 999–1002

[44]   Rahmstorf S 2002 Ocean circulation and climate during the past 120,000 years *Nature* **419** 207–14

[45] Caesar L, McCarthy G D, Thornalley D J R, Cahill N and Rahmstorf S 2021 Current Atlantic meridional overturning circulation weakest in last millennium *Nature Geoscience* **14** 118–20

[46] Agarwal A, Caesar L, Marwan N, Maheswaran R, Merz B and Kurths J 2019 Network-based identification and characterization of teleconnections on different scales *Sci Rep* **9** 8808

[47] Crutzen P and Stoermer E 2000 International Geosphere Biosphere Programme (IGBP) Newsletter, 41

[48] Rai R and Sahu C K 2020 Driven by data or derived through physics? a review of hybrid physics guided machine learning techniques with cyber-physical system (cps) focus *IEEE Access* **8** 71050–73

[49] Karpatne A, Atluri G, Faghmous J H, Steinbach M, Banerjee A, Ganguly A, Shekhar S, Samatova N and Kumar V 2017 Theory-guided data science: A new paradigm for scientific discovery from data *IEEE Transactions on knowledge and data engineering* **29** 2318–31

[50] Reichstein M, Camps-Valls G, Stevens B, Jung M, Denzler J and Carvalhais N 2019 Deep learning and process understanding for data-driven Earth system science *Nature* **566** 195–204

[51] Faghmous J H and Kumar V 2014 A big data guide to understanding climate change: The case for theory-guided data science *Big data* **2** 155–63

[52] Xu T and Valocchi A J 2015 Data-driven methods to improve baseflow prediction of a regional groundwater model *Computers & Geosciences* **85** 124–36

[53] Bozzolan E, Holcombe E, Pianosi F and Wagener T 2020 Including informal housing in slope stability analysis–an application to a data-scarce location in the humid tropics *Natural Hazards and Earth System Sciences* **20** 3161–77

[54] Hao Z, Singh V P and Hao F 2018 Compound extremes in hydroclimatology: a review *Water* **10** 718

[55] Zscheischler J, Martius O, Westra S, Bevacqua E, Raymond C, Horton R M, van den Hurk B, AghaKouchak A, Jézéquel A and Mahecha M D 2020 A typology of compound weather and climate events *Nature reviews earth & environment* **1** 333–47

[56] Raymond C, Horton R M, Zscheischler J, Martius O, AghaKouchak A, Balch J, Bowen S G, Camargo S J, Hess J and Kornhuber K 2020 Understanding and managing connected extreme events *Nature climate change* **10** 611–21

[57] Hao Z and Singh V P 2020 Compound events under global warming: a dependence perspective *Journal of Hydrologic Engineering* **25** 03120001